\documentclass[10pt]{IEEEtran}
\IEEEoverridecommandlockouts
\usepackage{cite}
\usepackage{amsmath,amssymb,amsfonts}
\usepackage{graphicx}
\usepackage{textcomp}
\usepackage{amsmath}
\usepackage{booktabs} 
\usepackage[skip=1pt, font=footnotesize]{caption}
\usepackage{multirow}
\usepackage{hyperref}       
\usepackage{url}            
\usepackage{setspace}
\usepackage{booktabs}       
\usepackage{amsfonts}       
\usepackage{nicefrac}       
\usepackage{microtype}      
\usepackage{xcolor}         
\usepackage{algorithm}
\usepackage{algpseudocode}
\usepackage{amsmath}
\usepackage{tikz}
\usepackage{quantikz}

\usepackage{orcidlink}
\usepackage{comment}
\usepackage{amssymb}
\usepackage{braket}
\usepackage{stfloats}
\usepackage{xcolor}
\def\BibTeX{{\rm B\kern-.05em{\sc i\kern-.025em b}\kern-.08em
    T\kern-.1667em\lower.7ex\hbox{E}\kern-.125emX}}

\usepackage{enumitem}
\setlist[itemize]{leftmargin=*}%
\setlist[enumerate]{leftmargin=*}%

\usepackage[compact]{titlesec}
\usepackage{stfloats}

\titlespacing\section{0pt}{0.2\baselineskip}{0.12\baselineskip}
\titlespacing\subsection{0pt}{0.15\baselineskip}{0.08\baselineskip}
\titlespacing\subsubsection{0pt}{0.1\baselineskip}{0.08\baselineskip}

\makeatletter
\def\bstctlcite{\@ifnextchar[{\@bstctlcite}{\@bstctlcite[\@IEEEstringA]}}
\def\@bstctlcite[#1]#2{\gdef\@IEEEcurrentbstctl{#2}}
\IEEEtriggercmd{\reset@font\normalfont\refsize} 
\IEEEtriggeratref{1} 
\makeatother

\usepackage{fancyhdr,lipsum}
\fancypagestyle{firstpage}{
  \fancyhf{}
  \fancyhead[C]{To appear at the IEEE International Conference on Quantum Artificial Intelligence (QAI), Nottingham, UK, December 2026.}
  \fancyfoot[C]{\thepage}
}

\begin{document}
\pagestyle{empty}
\bstctlcite{bstctl:etal, bstctl:nodash, bstctl:simpurl}

\title{Q-PhotoMarket: A Design Space Exploration Framework for Photonic Hybrid Quantum Neural Networks in Financial Market Prediction
}

\author{\IEEEauthorblockN{Alberto Marchisio\IEEEauthorrefmark{4}\IEEEauthorrefmark{2}\IEEEauthorrefmark{3}$^*$\thanks{$^*$This work has been carried out while at NYUAD.}, Hanzalah Mohamed Siraj\IEEEauthorrefmark{4}\IEEEauthorrefmark{2}, Muhammad Kashif\IEEEauthorrefmark{4}\IEEEauthorrefmark{2}, Nouhaila Innan\IEEEauthorrefmark{4}\IEEEauthorrefmark{2}, Muhammad Shafique\IEEEauthorrefmark{4}\IEEEauthorrefmark{2}}

\IEEEauthorblockA{\IEEEauthorrefmark{4} \normalsize eBrain Lab, Division of Engineering, New York University Abu Dhabi, PO Box 129188, Abu Dhabi, UAE\\}
\IEEEauthorblockA{\IEEEauthorrefmark{2} \normalsize Center for Quantum and Topological Systems, NYUAD Research Institute, New York University Abu Dhabi, UAE\\}
\IEEEauthorblockA{\IEEEauthorrefmark{3}
\normalsize SDU Microelectronics, Institute of Mechanical and Electrical Engineering, University of Southern Denmark, Odense, Denmark\\
Emails: chisio@sdu.dk, \{sm12152, muhammadkashif, nouhaila.innan, muhammad.shafique\}@nyu.edu}
\vspace{-25pt}
}

\thispagestyle{empty}
\maketitle
\thispagestyle{empty}
\pagenumbering{gobble}
\thispagestyle{firstpage}

\begin{spacing}{0.92}

\begin{small}
\begin{abstract}
Photonic quantum computing has recently emerged as a promising platform for hybrid quantum machine learning due to its native realization of linear-optical circuits and the computational complexity of boson sampling. However, despite growing interest in quantum methods for finance, the influence of photonic circuit design choices on predictive performance remains largely unexplored. Existing studies typically evaluate a single architecture, leaving the broader photonic design space unexamined.
In this work, we present \textbf{Q-PhotoMarket}, a systematic design space exploration (DSE) framework for photonic hybrid quantum neural networks (HQNNs) applied to financial market prediction. We explore over 5,000 valid photonic configurations spanning input photon states, circuit architectures, entangling models, and measurement strategies across their compatible computation spaces, for U.S., Indian, and cryptocurrency markets. To improve search efficiency, the exhaustive exploration is complemented with Bayesian optimization. We further incorporate threshold calibration and prediction-collapse diagnostics to enable reliable evaluation under increasingly imbalanced return thresholds.
Experimental results show that systematic exploration of more than 5,000 photonic HQNN configurations reveals consistent architectural patterns across financial markets, identifies robust high-performing designs, and demonstrates competitive performance relative to classical machine learning baselines.
\end{abstract}

\begin{IEEEkeywords}
Quantum Machine Learning, Photonic Quantum Computing, Design Space Exploration, Financial Market Prediction
\end{IEEEkeywords}
\end{small}
\vspace{-10pt}
\end{spacing}

\begin{spacing}{0.98}

\section{Introduction}

Quantum Machine Learning (QML) has emerged as a promising direction for extending machine learning with quantum computational primitives~\cite{cerezo2021variational, zaman2023survey}. Among the various QML paradigms, Hybrid Quantum Neural Networks (HQNNs) have attracted considerable attention due to their compatibility with near-term quantum hardware and their ability to combine quantum representations with classical optimization~\cite{cerezo2022challenges}. While HQNNs have demonstrated encouraging results across a growing range of applications, understanding how quantum architectural choices influence learning performance remains an important open problem.

Recent advances in photonic quantum computing have established linear-optical processors as a promising alternative to conventional gate-based quantum hardware. Compared with many qubit-based platforms, photonic systems offer practical advantages such as room-temperature operation, relatively low susceptibility to decoherence, and compatibility with existing optical communication technologies~\cite{zhangPhotonic2024, maring2023general}. Coupled with increasingly mature software ecosystems, these advances have made photonic HQNNs more accessible for developing and evaluating practical quantum machine learning models~\cite{heurtel2023perceval, notton2026merlin}. Unlike gate-based implementations, however, photonic HQNNs expose a rich architectural design space involving multiple interacting design decisions~\cite{maring2024versatile, elnakhal2026q}. Despite this flexibility, existing studies typically adopt a single manually designed architecture, leaving the impact of these design choices largely unexplored. This gap is particularly important given the demonstrated sensitivity of HQNN performance to circuit design in gate-based models~\cite{kashif2021design} and the computational potential of linear-optical circuits based on boson sampling~\cite{aaronson2011computational}.

\begin{figure}[t]
    \centering
    \includegraphics[width=0.98\columnwidth]{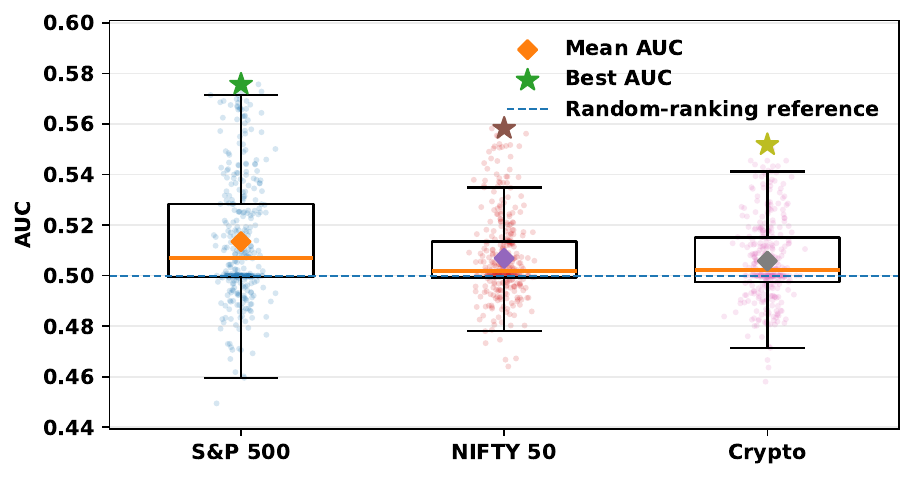}
    \caption{Distribution of AUC across valid photonic HQNN configurations for the S\&P~500, NIFTY~50, and cryptocurrency markets.}
    \label{fig:motivational_dse}
\end{figure}

To illustrate the practical importance of this design problem, Figure~\ref{fig:motivational_dse} shows the distribution of AUC across the explored photonic HQNN configurations for the three considered markets. Although the median performance remains close to the random-ranking reference, the observed distributions are broad, and the best configurations achieve notably higher AUC. This spread indicates that predictive performance is highly sensitive to photonic design choices and that manually selecting a single architecture may lead to misleading conclusions about the capability of a photonic HQNN. 

Financial market prediction provides a realistic and challenging benchmark for studying this problem. The task is characterized by weak predictive signals, non-stationary dynamics, and severe class imbalance, while state-of-the-art classical approaches generally achieve only modest improvements under rigorous walk-forward evaluation~\cite{Ayyildiz2024}. Existing QML studies for finance typically employ fixed quantum circuit architectures with limited architectural analysis~\cite{Innan2024, choudhary2026hqnn}, making it difficult to understand how photonic design choices influence predictive performance.

To address this gap, we propose \textbf{Q-PhotoMarket}, a systematic design space exploration (DSE) framework for photonic HQNNs. Q-PhotoMarket explores multiple photonic design dimensions, including input photon states, circuit architectures, entangling models, measurement strategies, and computation spaces, while combining exhaustive exploration with Bayesian optimization to efficiently identify promising regions of the design space.

The main contributions of this work are:

\begin{itemize}
    \item We propose \textbf{Q-PhotoMarket}, a systematic DSE framework for photonic HQNNs, enabling the analysis of key photonic design dimensions for real-world QML applications.

    \item We perform a large-scale empirical evaluation of the photonic design space by exploring thousands of valid HQNN configurations across multiple financial markets and return prediction thresholds.

    \item We integrate exhaustive exploration with Bayesian optimization to efficiently identify promising photonic architectures while reducing the search cost of high-dimensional design spaces.

    \item We provide empirical insights into the impact of photonic architectural choices on HQNN performance and establish a reproducible benchmark to support future research in photonic QML.
\end{itemize}

\section{Background and Related Work}

\subsection{Photonic HQNNs}
HQNNs integrate parameterized quantum circuits with classical neural networks, combining quantum feature transformations with classical optimization. In photonic HQNNs, information is encoded into bosonic modes and processed through trainable linear-optical circuits composed of beam splitters, phase shifters, and interferometers. Universal interferometer decompositions, such as the Reck and Clements architectures, enable the realization of arbitrary unitary transformations and constitute the foundation of many photonic quantum learning models~\cite{clements2016optimal}. Recent software frameworks, including Perceval and Merlin, have further facilitated the development of photonic QML models by providing differentiable simulation environments for training and evaluation~\cite{heurtel2023perceval, notton2026merlin}.

\subsection{Design Space Exploration in HQNNs}
The predictive performance of HQNNs is strongly influenced by architectural design choices, including data encoding, circuit topology, variational ans\"atze, measurement strategies, and optimization protocols. Consequently, recent studies have explored systematic design space exploration (DSE) for gate-based HQNNs to identify architectures that balance expressivity, trainability, and hardware efficiency~\cite{kashif2021design, du2022quantum, he2024training}. In contrast, comparable studies for photonic HQNNs remain scarce, despite the considerably richer design space introduced by photon-state preparation, linear-optical circuit topologies, and measurement schemes~\cite{maring2024versatile}. Existing photonic QML studies typically evaluate a single manually designed architecture, providing limited understanding of how individual design choices influence learning performance. This gap motivates the systematic photonic DSE framework proposed in this work.

\section{Q-PhotoMarket Framework}

\begin{figure*}[t!]
    \centering
    \includegraphics[width=1.0\linewidth]{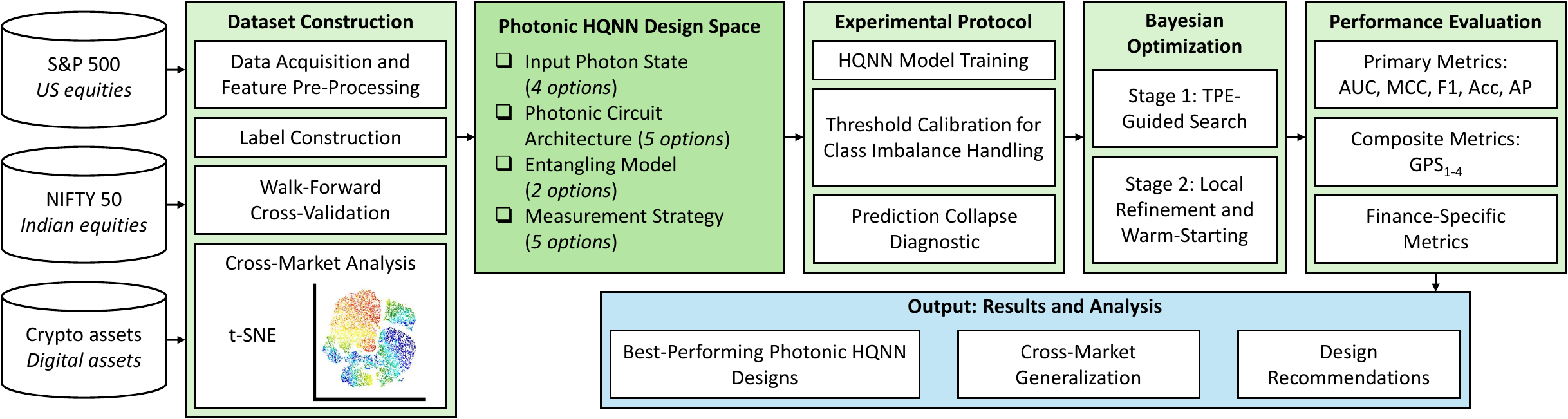}
    \caption{Overview of the proposed Q-PhotoMarket framework.}
    \label{fig:framework}
\end{figure*}

Q-PhotoMarket systematically explores the design space of photonic hybrid quantum neural networks (HQNNs) for financial market prediction across three asset classes: U.S. equities (S\&P~500), Indian equities (NIFTY~50), and cryptocurrencies. The design space comprises four independent dimensions: input photon state, photonic circuit architecture, entangling model, and measurement strategy. The corresponding computation space is determined by the selected measurement strategy. As illustrated in Figure~\ref{fig:framework}, the framework consists of four main stages: (i) leakage-safe dataset construction, (ii) photonic HQNN design space exploration, (iii) unified experimental evaluation, and (iv) Bayesian optimization for refining promising configurations. 

\subsection{Problem Formulation}

We formulate financial market prediction as a binary classification task that predicts the direction of the next-day asset return. Let $r_{t \rightarrow t+h}$ denote the forward return of an asset over prediction horizon $h$ from time $t$. The binary label is defined as
\begin{equation}
    y_t = \mathbf{1}\!\left(r_{t\rightarrow t+h} > \tau\right),
    \label{eq:label}
\end{equation}

where $\mathbf{1}(\cdot)$ is the indicator function and $\tau$ is a user-defined return threshold. Setting $\tau=0$ corresponds to conventional up/down direction prediction, whereas larger thresholds ($\tau \in \{0.5\%,1.0\%\}$) require economically more significant price movements, resulting in progressively imbalanced classification tasks. Throughout this work, we evaluate three thresholds, $\tau \in \{0,0.5\%,1.0\%\}$, with a one-day prediction horizon ($h=1$).

To ensure realistic evaluation on non-stationary financial time series, all experiments employ an expanding-window walk-forward protocol where each validation window strictly follows its corresponding training window to prevent look-ahead bias.

\subsection{Dataset Construction}

\subsubsection{Data Acquisition and Feature Engineering}
Daily OHLCV (Open, High, Low, Close, Volume) data for each asset class are collected from Yahoo Finance for the period 2018-2025. For each asset, we construct a leakage-safe feature set using only information available at time $t$. The resulting dataset comprises 23 engineered features covering return and momentum, volatility and risk, technical indicators, liquidity and microstructure, and price structure:

\begin{itemize}
  \item \textbf{Return and momentum}: \texttt{ret\_1}, \texttt{ret\_5}, \texttt{ret\_10}, \texttt{ret\_20}, \texttt{cross\_momentum}
  \item \textbf{Volatility and risk}: \texttt{vol\_5}, \texttt{vol\_10}, \texttt{vol\_20}, \texttt{vol\_ratio\_5\_20}, \texttt{ret\_autocorr\_20}
  \item \textbf{Technical indicators}: \texttt{rsi\_14}, \texttt{sma20\_dist}, \texttt{bb\_width\_20}, \texttt{vwap\_deviation}
  \item \textbf{Liquidity and microstructure}: \texttt{volume\_ratio\_20}, \texttt{amihud\_illiquidity}, \texttt{obv\_slope\_10}
  \item \textbf{Price structure and event}: \texttt{high\_low\_ratio}, \texttt{overnight\_gap}, \texttt{day\_of\_week}, \texttt{vix\_bucket}, \texttt{month\_end\_flag}, \texttt{earnings\_event\_flag}
\end{itemize}

Macro-contextual features are incorporated through a 5-day rolling index return (\texttt{index\_ret\_5d}), a discretized VIX volatility regime proxy (\texttt{vix\_bucket}), and a binary earnings-event indicator (\texttt{earnings\_event\_flag}). Samples with missing values are removed before label construction.

To satisfy the photonic mode constraints imposed by the Merlin simulation framework, the HQNN operates on a fixed subset of 16 input features throughout all experiments. The selected features correspond exclusively to price-derived market descriptors, including return, momentum, volatility, technical, liquidity, and price-structure indicators, while external contextual signals (e.g., market-index and event-related features) are excluded. 

\subsubsection{Label Construction}
Future return labels are generated only after feature engineering to eliminate look-ahead leakage. For each threshold $\tau$, labels are computed independently using the forward return $r_{t\rightarrow t+h}=P_{t+h}/P_t-1$. As $\tau$ increases, the positive class prevalence decreases from approximately 50\% ($\tau=0$) to 20--34\% ($\tau=1\%$), inducing increasing label imbalance across all three markets.

\subsubsection{Walk-Forward Cross-Validation}
To respect the temporal ordering of financial data and prevent look-ahead bias, we construct time-consistent evaluation folds using an expanding-window walk-forward scheme. 
Training windows grow monotonically, while each validation window is strictly subsequent to its corresponding training window ($n_{\mathrm{folds}}=5$, minimum training fraction $=0.5$). This protocol provides reproducible evaluation across all dataset-threshold combinations.

\subsubsection{Dataset and Cross-Market Feature-Space Analysis}
The proposed framework is evaluated on three market domains: U.S. equities, Indian equities, and cryptocurrencies. Daily OHLCV data from 2018-2025 are transformed into a common leakage-safe feature representation and subsequently analyzed across markets. Since these datasets exhibit different volatility characteristics, trend persistence, and class distributions, we assess the similarity of their engineered feature spaces before performing cross-market architectural comparisons.

To this end, we employ t-distributed Stochastic Neighbor Embedding (t-SNE)~\cite{van2008visualizing} as a qualitative visualization technique. Feature vectors are first normalized to $[0,1]$, pre-reduced using PCA for embedding stability, and projected into two dimensions using a perplexity of $\min(30,N/4)$, 500 optimization iterations, and a fixed random seed.

\begin{figure}[t!]
  \centering
  \includegraphics[width=0.9\columnwidth]{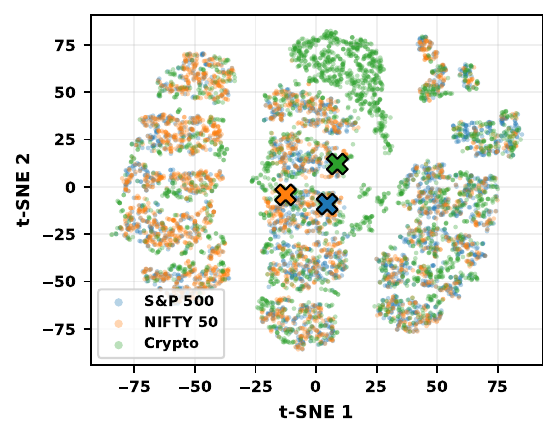}
  \caption{Two-dimensional t-SNE visualization of the engineered feature spaces for the S\&P~500, NIFTY~50, and cryptocurrency datasets.}
  \label{fig:tsne_markets}
\end{figure}

As shown in Figure~\ref{fig:tsne_markets}, the three markets occupy overlapping yet distinguishable regions in feature space. This qualitative analysis suggests that the datasets share sufficient statistical structure to support meaningful cross-market comparisons while preserving market-specific characteristics, allowing architectural differences to be interpreted with reduced influence from dataset-level distributional effects.

\subsection{Photonic HQNN Architecture and Design Space}

The proposed HQNN is implemented using the Merlin photonic simulation framework and trained end-to-end in PyTorch. Since Merlin provides a differentiable simulation backend, gradients are computed through standard backpropagation without requiring parameter-shift evaluation. The hybrid architecture couples a trainable photonic \texttt{QuantumLayer} with a classical MLP prediction head (\texttt{LazyLinear}$\to$\texttt{ReLU}$\to$\texttt{Linear(1)}), which maps measurement-derived photonic features to the final prediction logit. Throughout the design space exploration, the number of photonic modes is fixed as $n_{\mathrm{modes}} = n_{\mathrm{features}} + 1$, with $n_{\mathrm{features}}=16$. 
Q-PhotoMarket explores four configurable dimensions of the photonic HQNN architecture: input photon state, circuit architecture, entangling model, and measurement strategy. The resulting design space is summarized in Table~\ref{tab:dse}.

\begin{table}[t]
\centering
\caption{Photonic HQNN design space explored by Q-PhotoMarket. Computation space is determined by the selected measurement strategy. Total valid configurations per (market, $\tau$) are approximately 600 after invalid-combination pruning.}
\label{tab:dse}
\resizebox{\columnwidth}{!}{%
\begin{tabular}{lll}
\toprule
\textbf{Dimension} & \textbf{Options} & \textbf{Count} \\
\midrule
Input photon state   & Alternating, Bunched, Pair-bunched, Thermal-like & 4 \\
Circuit architecture & Simple, Ring NN, Clements Mesh, Phase-only,  & 5 \\
& Permuted Ring \\
Entangling model     & MZI, Bell & 2 \\
Measurement strategy & Probs-Fock, Probs-Unbunched, Probs-Dual-Rail,          & 5 \\
                     & Mode expectations, Mode correlations                    &   \\
\midrule
\multicolumn{2}{l}{\textbf{Valid configs per (market, $\tau$), after pruning}} & $\approx$\textbf{600} \\
\multicolumn{2}{l}{\textbf{Total runs ($\times$3 markets, $\times$3 thresholds)}} & $\approx$\textbf{5{,}400} \\
\bottomrule
\end{tabular}
}
\end{table}




\subsubsection{Input Photon State}
The input state determines the initial photon occupation used to encode classical features into the photonic Hilbert space. We consider four configurations:
\begin{itemize}
    \item \textbf{Alternating}: photons distributed across alternating modes.
    \item \textbf{Bunched}: photons initialized in adjacent low-index modes.
    \item \textbf{Pair-bunched}: photons injected in paired modes.
    \item \textbf{Thermal-like}: photon occupations sampled from a thermal distribution.
\end{itemize}

\subsubsection{Photonic Circuit Architecture}
Five linear-optical circuit topologies are investigated:
\begin{itemize}
    \item \textbf{Simple}: single-layer beamsplitter mesh.
    \item \textbf{Ring NN}: circular beamsplitter topology.
    \item \textbf{Clements Mesh}: universal rectangular interferometer~\cite{clements2016optimal}, providing maximal expressivity.
    \item \textbf{Phase-only}: phase-shifter network without beamsplitter entanglement.
    \item \textbf{Permuted Ring}: ring topology with randomized mode permutations.
\end{itemize}

\subsubsection{Entangling Model}
Two entangling primitives are evaluated:
\begin{itemize}
  \item \textbf{MZI}: Mach-Zehnder Interferometer, the canonical two-mode entangling unit.
  \item \textbf{Bell}: Bell-state-inspired entangling structure, generating stronger inter-mode correlations.
\end{itemize}

\subsubsection{Measurement Strategy}
Five measurement strategies are explored to extract classical representations from the output photonic state. Invalid combinations (e.g., non-binary input states paired with
\texttt{probs\_unbunched} or \texttt{probs\_dual\_rail}) are discarded
during the search. The associated computation space is determined by the selected measurement strategy. 

\begin{itemize}
    \item \textbf{Fock probabilities} (\texttt{probs\_fock}): output occupation probabilities in the Fock basis; restricted to $n_{\mathrm{modes}}\le4$ because of memory requirements.
    \item \textbf{Unbunched probabilities} (\texttt{probs\_unbunched}): distinguishable-photon probabilities in the unbunched computation space.
    \item \textbf{Dual-rail probabilities} (\texttt{probs\_dual\_rail}): probabilities obtained under dual-rail logical encoding.
    \item \textbf{Mode expectations} (\texttt{mode\_expectations}): first-order expectations $\langle\hat{n}_i\rangle$.
    \item \textbf{Mode correlations} (\texttt{mode\_correlations}): second-order correlations $\langle\hat{n}_i\hat{n}_j\rangle$.
\end{itemize}

\subsection{Experimental Protocol}

\subsubsection{HQNN Model Training}
All HQNN configurations are trained under a unified protocol. 
We employ the Adam optimizer with learning rate $\eta=3\times10^{-4}$ and binary cross-entropy loss (\texttt{BCEWithLogitsLoss}). Each model is trained for 8 epochs during Stage~1 and for 20 epochs during Stage~2, using a batch size of 128 and 100 measurement shots per circuit execution with a random seed fixed to 42. An optional cosine annealing scheduler and early stopping based on validation AUC (patience = 5 epochs) are supported. Gradients are propagated through Merlin's differentiable simulation
backend using automatic differentiation~\cite{notton2026merlin}.


\subsubsection{Class Imbalance Handling and Threshold Calibration}

Increasing the return threshold $\tau$ substantially reduces the positive-class prevalence, producing increasingly imbalanced classification tasks. For $\tau>0$, class imbalance is mitigated through inverse-frequency class weighting,

\begin{equation}
w_{+}=\frac{N_{-}}{\max(N_{+},1)},
\label{eq:classweight}
\end{equation}

where $N_{+}$ and $N_{-}$ denote the numbers of positive and negative training samples, respectively. Empirically, S\&P~500 and NIFTY~50 achieve the best performance using \texttt{class-weighting=balanced}, whereas cryptocurrency datasets benefit from \texttt{class-weighting=none}.

To avoid suboptimal fixed decision boundaries, the classification threshold is calibrated on the validation fold by maximizing a selected objective using Eq.~\ref{eq:thresh_cal}, where $\text{Obj} \in \{\text{MCC},\, \text{F1},\, \text{BalAcc}\}$.

\begin{equation}
  \hat{\tau}^* = \arg\max_{\tau \in [0,1]} \;\text{Obj}\!\left(y_{\text{val}},\, \mathbf{1}(\hat{p}_{\text{val}} \geq \tau)\right),
  \label{eq:thresh_cal}
\end{equation}

\subsubsection{Prediction Collapse Diagnostics}
\label{sec:collapse}
A common failure mode was identified at $\tau > 0$, where positive class prevalence drops to 20-34\% and predicted probabilities remain below the conventional decision threshold of 0.5, causing all samples to be assigned to the majority class 0. Although ranking performance (AUC) may remain meaningful, threshold-dependent metrics such as F1-score and MCC collapse to zero,


\begin{equation}
  \text{F1} = 0, \quad \text{MCC} = 0, \quad \text{BalAcc} = 0.5,
  \label{eq:collapse}
\end{equation}


To identify this behavior, each experiment records the calibrated decision threshold, its optimization objective, the positive prediction rate (PPR), and a prediction-collapse warning when $\mathrm{PPR}\notin[0.02,0.98]$.

\begin{figure}[t]
  \centering
  \includegraphics[width=0.98\columnwidth]{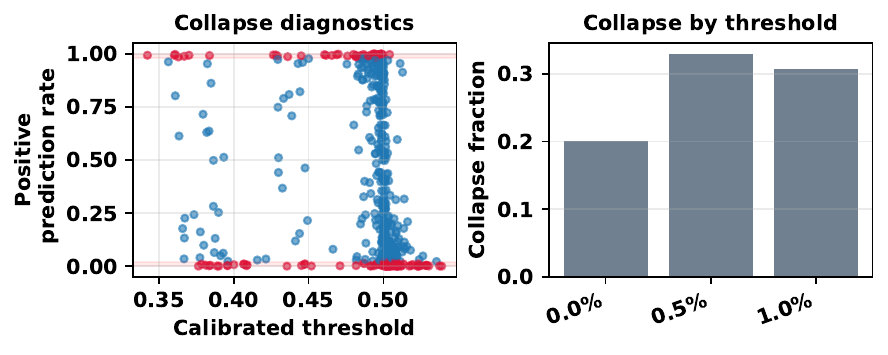}
  \caption{Prediction-collapse diagnostics. Left: calibrated decision thresholds versus positive prediction rate (PPR). Right: fraction of configurations exhibiting prediction collapse under different return thresholds.}
  \label{fig:collapse_diagnostics}
\end{figure}

Figure~\ref{fig:collapse_diagnostics} illustrates the effectiveness of this diagnostic procedure. The left panel shows that calibrated thresholds successfully recover meaningful positive prediction rates across a wide range of operating points, while the right panel quantifies how the frequency of prediction collapse increases as the return threshold becomes more stringent. These diagnostics provide an additional validation of the proposed calibration strategy before comparing HQNN architectures.

\subsection{Bayesian Optimization}

To complement the exhaustive grid search, we employ Bayesian optimization (BO) using the Tree-structured Parzen Estimator (TPE) sampler implemented in Optuna~\cite{akiba2019optuna}. BO explores the same photonic design space while efficiently concentrating the search on promising regions identified during earlier trials.

\subsubsection{Stage 1: TPE-Guided Search}
The TPE sampler models the distribution of high-performing configurations and proposes new candidates by jointly optimizing the photonic design parameters, the number of input features, and the circuit shot count. Candidate configurations are ranked using the composite objective
\begin{equation}
  \mathcal{L} = w_{\text{AUC}} \cdot \text{AUC} + w_{\text{F1}} \cdot \text{F1} + w_{\text{GPS}} \cdot \text{GPS} - \delta_{\text{collapse}},
  \label{eq:bo_obj}
\end{equation}
where $w_{\text{AUC}} = 0.6$, $w_{\text{F1}} = 0.3$, $w_{\text{GPS}} = 0.1$, and $\delta_{\text{collapse}} = 0.15$ is a penalty applied when $\text{PPR} \notin [0.05,\, 0.95]$. A resource guardrail skips configurations whose heuristic cost index, defined as $(n_{\text{modes}}^3 \times m_{\text{factor}} \times a_{\text{factor}} \times \text{shots}/100)$, exceeds a threshold, preventing memory or runtime instability for high-mode, high-shot combinations. A median pruner with $t_{\text{warmup}} = 5$ epochs additionally terminates unpromising trials early.

\subsubsection{Stage 2: Local Refinement and Warm-Starting}
After Stage~1, the top-$k$ discovered configurations are used to seed a local dense refinement pass. In subsequent studies, prior sweep results 
are enqueued directly into the optimization process, 
allowing the sampler to incorporate all prior grid and confirm-run evidence before proposing new trials. Warm-start rows are ranked by AUC and filtered to respect mode and shot constraints.

\subsubsection{Optimization Protocol}
The BO phase is performed independently for each market with separate Optuna studies, using $n_{\text{trials}} = 150$ per shard, $n_{\text{startup}} = 30$ random trials before activating the TPE sampler, and a per-trial subprocess timeout of 300 seconds. The resulting studies are merged for downstream analysis, enabling parameter-importance estimation and convergence analysis across the explored photonic design space.


\subsection{Performance Evaluation}

\subsubsection{Primary Metrics}

Increasing return thresholds produce progressively imbalanced classification tasks, making raw accuracy alone insufficient for performance evaluation. Accordingly, we emphasize threshold-robust metrics including the Area Under the ROC Curve (AUC), Matthews Correlation Coefficient (MCC), and balanced accuracy, while reporting conventional accuracy only as a complementary measure.
All configurations are evaluated using accuracy, F1-score, MCC, Area Under the ROC Curve (AUC), and Average Precision (AP). MCC is defined as:

\begin{small}
\begin{equation}
  \text{MCC} = \frac{TP \cdot TN - FP \cdot FN}{\sqrt{(TP{+}FP)(TP{+}FN)(TN{+}FP)(TN{+}FN)}},
  \label{eq:mcc}
\end{equation}
\end{small}

and is particularly suitable for imbalanced financial classification because it accounts for all elements of the confusion matrix. Balanced accuracy is defined as $(TPR+TNR)/2$.

\subsubsection{Composite Metrics}
Following~\cite{gps}, we report four General Performance Score (GPS) variants, each defined as the harmonic mean of a complementary subset of metrics:
\begin{small}
\begin{align}
  \text{GPS}_1 &= \frac{3}{\dfrac{1}{\text{BalAcc}} + \dfrac{1}{\text{AUC}} + \dfrac{1}{\text{F1}}} \label{eq:gps1}\\[6pt]
  \text{GPS}_2 &= \frac{4}{\dfrac{1}{\text{Acc}} + \dfrac{1}{\text{Prec}} + \dfrac{1}{\text{Recall}} + \dfrac{1}{\text{F1}}} \label{eq:gps2}\\[6pt]
  \text{GPS}_3 &= \frac{3}{\dfrac{1}{\text{BalAcc}} + \dfrac{1}{\text{AUC}} + \dfrac{1}{\overline{\text{Acc}}_\text{CV}}} \label{eq:gps3}\\[6pt]
  \text{GPS}_4 &= \frac{4}{\dfrac{1}{\text{BalAcc}} + \dfrac{1}{\text{AUC}} + \dfrac{1}{\text{F1}} + \dfrac{1}{\overline{\text{Acc}}_\text{CV}}} \label{eq:gps4}
\end{align}
\end{small}
where $\overline{\text{Acc}}_\text{CV}$ denotes the mean walk-forward cross-validation accuracy. These composite metrics provide a unified ranking while emphasizing different performance trade-offs.

\subsubsection{Finance-Specific Metrics}
To complement conventional classification metrics, we report two finance-oriented measures. \emph{Pseudo-Sharpe} approximates the risk-adjusted return of a strategy that enters a long position whenever $\hat{y}_t=1$, computed as the ratio between the mean predicted-positive return and its standard deviation. \emph{Magnitude-weighted accuracy} weights each correct prediction by $|r_{t\rightarrow t+h}|$, assigning greater importance to larger market movements.

\subsubsection{Baselines and Diagnostic Outputs}
For each dataset-threshold combination, we compare the proposed photonic HQNNs against three classical baselines: a majority-class classifier, logistic regression, and a shallow multilayer perceptron. These baselines provide reference points for assessing whether photonic quantum models offer meaningful improvements over classical machine learning approaches~\cite{kashif2021design}.

Each experiment records calibrated decision thresholds, positive prediction rates, prediction probabilities, and collapse diagnostics. Bayesian optimization studies additionally store the composite objective value, parameter importance estimates, and convergence history for subsequent analysis.


\section{Results and Discussion}

\subsection{Overall Design Space Exploration}

We first evaluate the complete photonic design space using the exhaustive screening protocol comprising approximately 1,080 configurations per market. For each market, approximately 600 candidate architectures are subsequently re-evaluated using multi-fold validation to confirm robustness. Throughout this section, we emphasize AUC, MCC, and balanced accuracy because increasing return thresholds introduce progressively stronger class imbalance.

\begin{figure}[t]
  \centering
  \includegraphics[width=0.98\columnwidth]{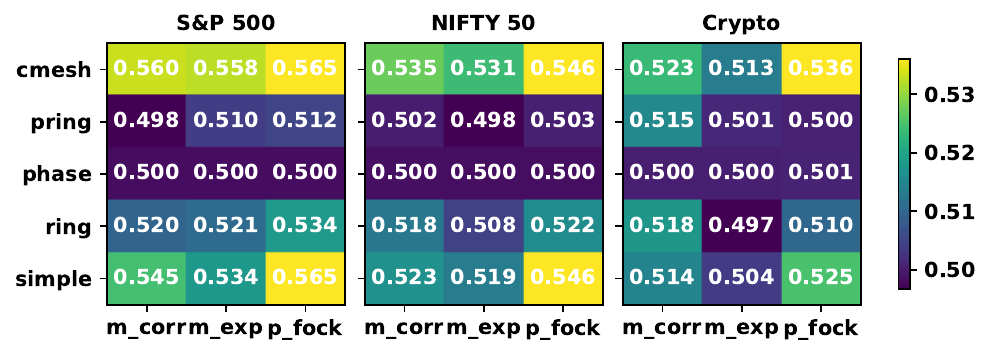}
  \caption{Mean AUC obtained across the explored photonic HQNN design space for different combinations of circuit architectures and measurement strategies at $\tau=1\%$.}
  \label{fig:auc_heatmap_arch_meas}
\end{figure}

Figure~\ref{fig:auc_heatmap_arch_meas} summarizes the mean AUC obtained for different combinations of photonic circuit architectures and measurement strategies under the most challenging return threshold ($\tau=1\%$). Several consistent trends emerge across all three markets. Architectures based on the \textit{Clements Mesh} and \textit{Simple} interferometer generally achieve higher mean AUC than the \textit{Phase-only} and \textit{Permuted Ring} variants, while probability-based measurements in the Fock basis frequently outperform expectation- and correlation-based alternatives. Although the optimal configuration remains market-dependent, these results demonstrate that architectural decisions have a measurable and systematic impact on HQNN performance.

\subsection{Influence of Photonic Design Choices}

\begin{figure*}[t]
  \centering
  \includegraphics[width=0.96\textwidth]{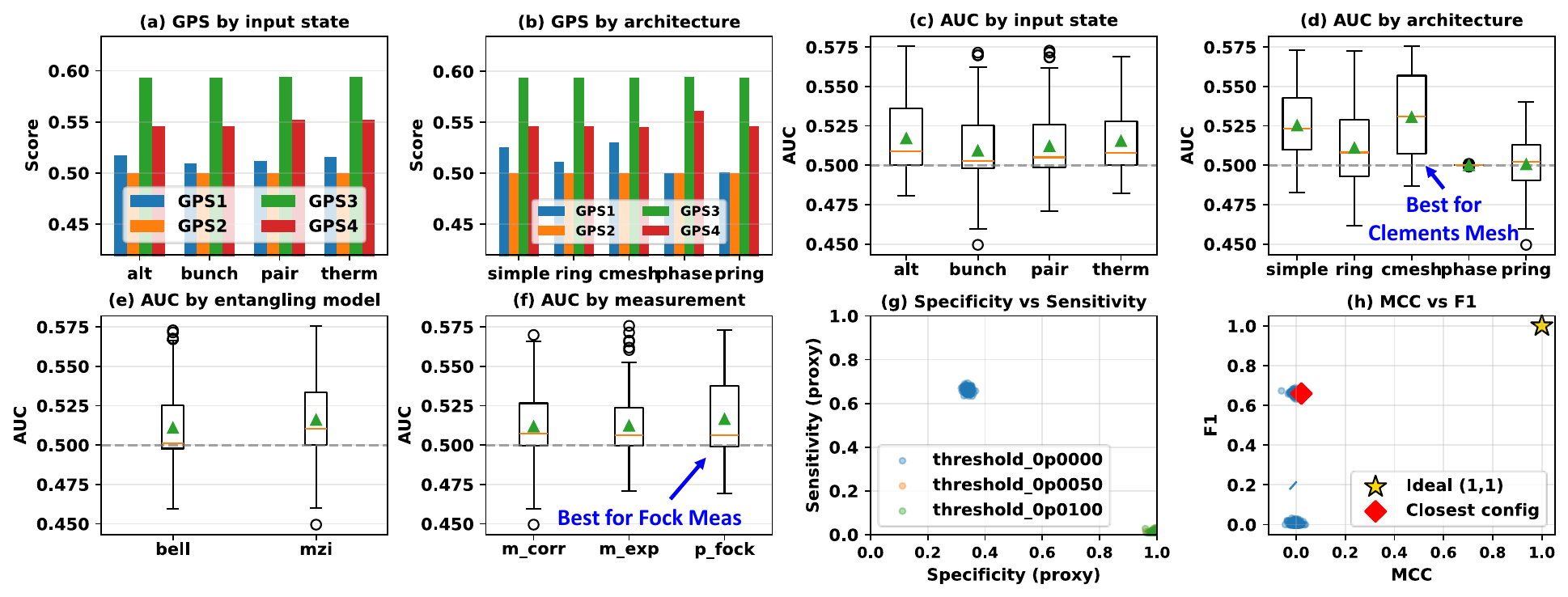}
  \caption{S\&P 500: 8-panel DSE diagnostics from the screening grid.}
  \label{fig:dse_sp500}
\end{figure*}

\begin{figure*}[t]
  \centering
  \includegraphics[width=0.96\textwidth]{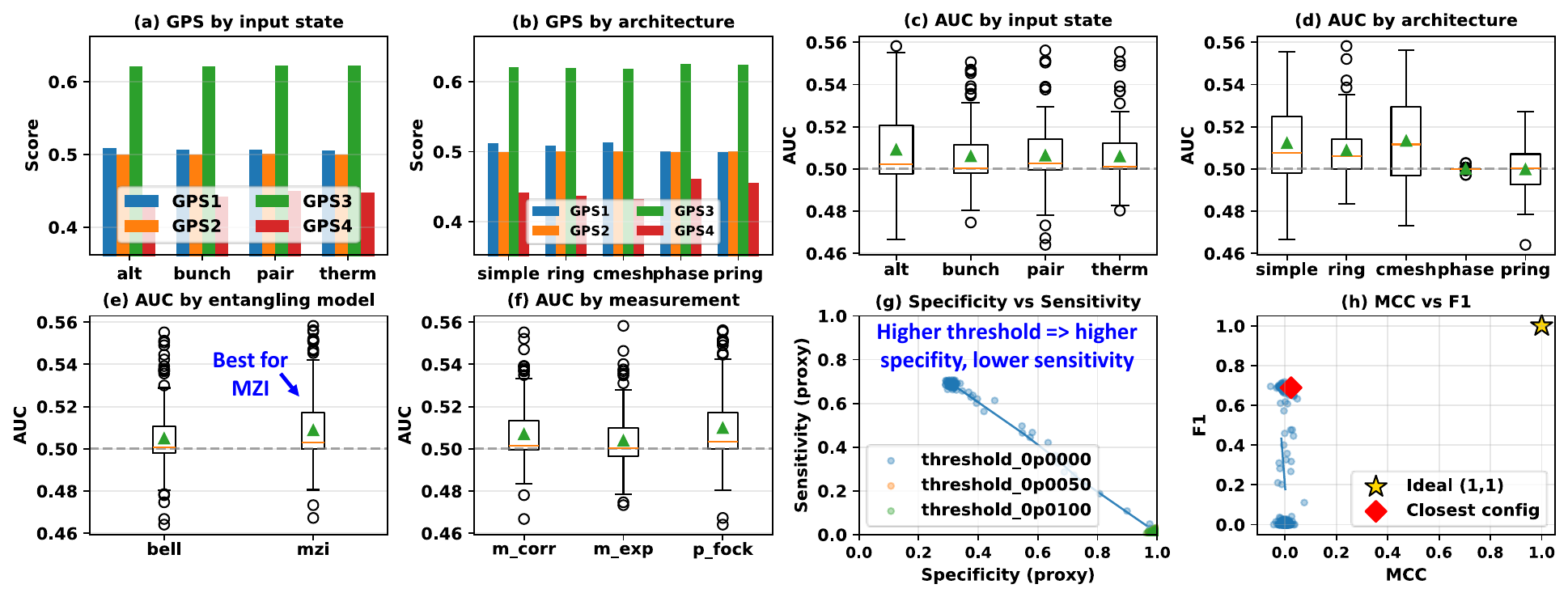}
  \caption{NIFTY 50: 8-panel DSE diagnostics from the screening grid.}
  \label{fig:dse_nifty50}
\end{figure*}

\begin{figure*}[t]
  \centering
  \includegraphics[width=0.96\textwidth]{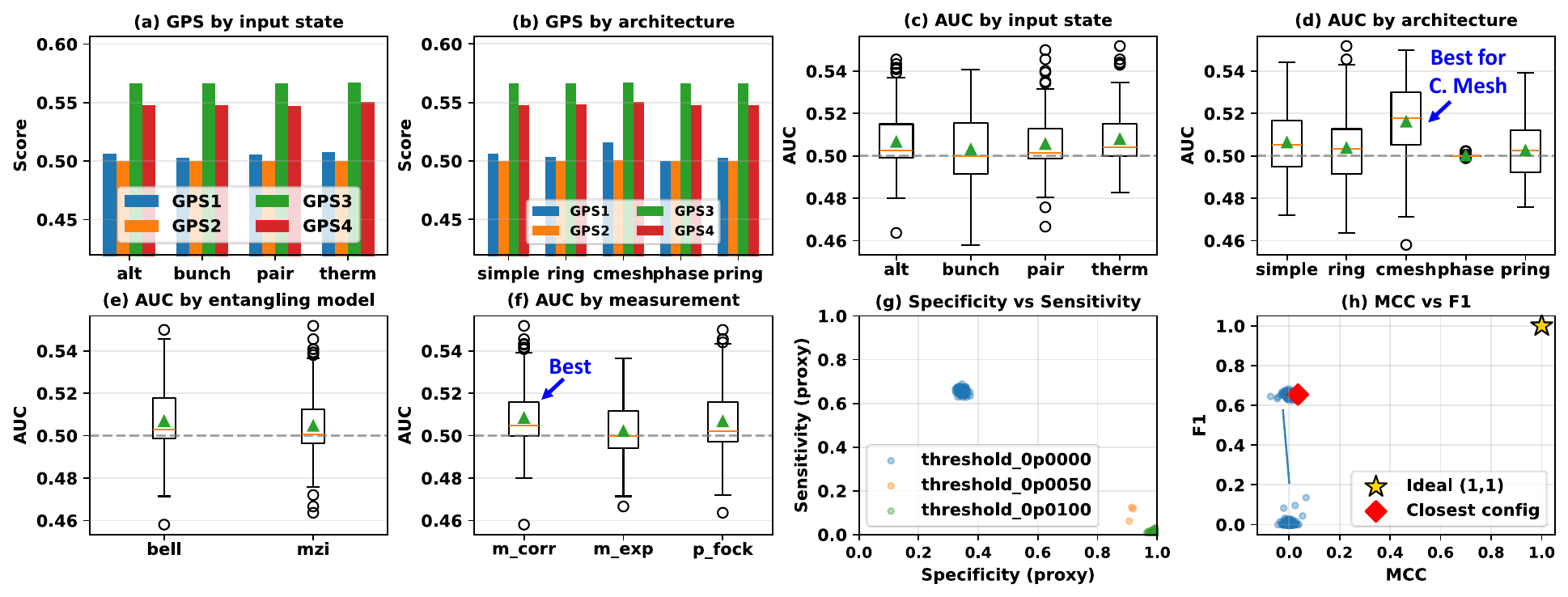}
  \caption{Crypto: 8-panel DSE diagnostics from the screening grid.}
  \label{fig:dse_crypto}
\end{figure*}

Figures~\ref{fig:dse_sp500}-\ref{fig:dse_crypto} illustrate the impact of the explored photonic design dimensions on HQNN performance across the three financial markets. The results reveal consistent performance differences arising from input photon states, circuit architectures, entangling models, and measurement strategies.

\textbf{Input photon state.}
The choice of input photon state has a relatively modest influence on predictive performance compared with the remaining architectural dimensions. Across all three markets, the four investigated input states exhibit similar GPS and AUC distributions, suggesting that photonic circuit design and measurement strategy play a more dominant role than the initial photon occupation.

\textbf{Circuit architecture.}
Among the investigated architectures, the \emph{Clements Mesh} consistently achieves the strongest average performance across markets, particularly for the S\&P~500 and cryptocurrency datasets. Ring-based architectures also demonstrate competitive behavior, whereas the \emph{Phase-only} architecture generally produces the weakest results. These observations indicate that expressive interferometer topologies provide a measurable advantage for financial prediction tasks.

\textbf{Entangling model.}
The choice of entangling primitive has a comparatively smaller impact than circuit topology. Across all three markets, the performance distributions of the Bell-inspired and Mach-Zehnder Interferometer (MZI) entangling models overlap substantially, although MZI exhibits a slight advantage on the NIFTY~50 dataset. Overall, these results suggest that circuit architecture contributes more strongly to predictive performance than the specific entangling primitive.

\textbf{Measurement strategy.}
Measurement strategy emerges as one of the most influential design dimensions. While probability-based Fock measurements provide strong performance on the S\&P~500 and NIFTY~50 datasets, correlation-based measurements become increasingly competitive on the crypto dataset. This behavior indicates that the optimal measurement representation is application-dependent and should be treated as an important hyperparameter during photonic HQNN design.

\subsection{Cross-Market and Threshold Sensitivity}

Having established the influence of individual photonic design dimensions, we next examine whether these trends remain consistent across financial markets and prediction settings.

\begin{figure}[t]
  \centering
  \includegraphics[width=0.98\columnwidth]{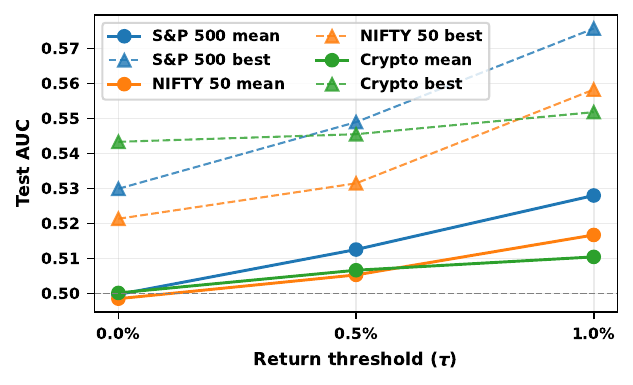}
  \caption{Threshold robustness of the explored photonic HQNN design space. Mean and best AUC are reported across all evaluated configurations, across $\tau\in\{0,0.5\%,1\%\}$.}
  \label{fig:threshold_robustness}
\end{figure}

Figure~\ref{fig:threshold_robustness} shows the mean and best AUC obtained across all evaluated configurations under increasing return thresholds. Across the three markets, both the average and best-performing photonic HQNNs exhibit improved ranking performance as the return threshold increases. The largest improvement is observed for the S\&P~500, where the mean AUC increases from 0.500 at $\tau=0$ to 0.528 at $\tau=1\%$, followed by NIFTY~50 (0.499$\rightarrow$0.517) and cryptocurrencies (0.500$\rightarrow$0.510). Although higher thresholds produce increasingly imbalanced datasets, they also define more discriminative prediction tasks by focusing on economically significant price movements, enabling improved class separability.

\subsection{Configuration Selection and Bayesian Refinement}

\begin{figure}[t]
  \centering
  \includegraphics[width=0.98\columnwidth]{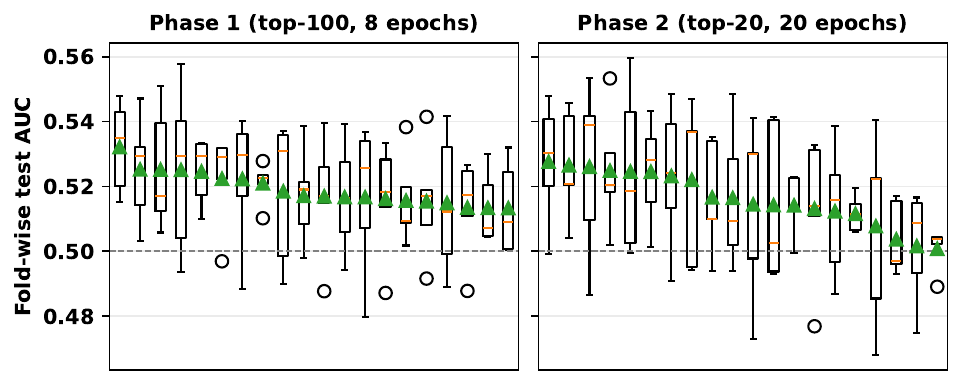}
  \caption{Fold-variance analysis for shortlisted configurations (Phases 1 vs 2).}
  \label{fig:fold_variance}
\end{figure}

Following the exhaustive screening stage, the highest-performing configurations were re-evaluated using the staged multi-fold protocol described in Section~III. Figure~\ref{fig:fold_variance} compares the fold-wise AUC distributions obtained during Phase~1 and Phase~2. While the variance across validation folds remains comparable, the longer Phase~2 training consistently improves the median AUC, indicating that the shortlisted architectures maintain their performance under more rigorous evaluation.

The highest-performing configurations obtained after Phase~2 are summarized for each market. For the S\&P~500, the best-performing architecture combines the \emph{Simple} circuit with Bell entangling and Fock probability measurements, achieving a mean AUC of 0.528 ($\pm$0.019). For NIFTY~50, the strongest configuration employs a \emph{Clements Mesh} architecture with MZI entangling and Fock measurements, reaching a mean AUC of 0.514 ($\pm$0.028).

Tables~\ref{tab:top10_overlap_sp500}-\ref{tab:top10_overlap_crypto} summarize representative high-performing configurations (top-10 per market) identified during the exhaustive search, revealing recurring architectural patterns across multiple evaluation metrics. In particular, Clements Mesh, Ring NN, and Simple architectures appear repeatedly among the highest-ranked configurations, confirming the trends observed during the design space exploration.

\begin{table*}[t]
\centering
\caption{Top-10 overlap table for S\&P 500. Top10Count indicates in how many of 8 metrics a configuration appears in the top-10.}
\label{tab:top10_overlap_sp500}
\setlength{\tabcolsep}{3pt}
\resizebox{.95\textwidth}{!}{%
\begin{tabular}{c l l l l l c l c c c}
\toprule
R & $\tau$ & Arch & Ent & Meas & State & Top10Count & MetricsHit & BestAUC & BestMCC & BestBalAcc \\
\midrule
1 & 0.0100 & simple & bell & probs\_fock & alternating & 4/8 & AUC, MCC, BalAcc, GPS & 0.573 & 0.000 & 0.500 \\
2 & 0.0100 & clements\_mesh & mzi & mode\_correlations & pair\_bunched & 4/8 & MCC, BalAcc, MWA, PSharpe & 0.557 & 0.008 & 0.501 \\
3 & 0.0000 & simple & mzi & mode\_expectations & thermal\_like & 4/8 & MCC, F1, BalAcc, AP & 0.519 & 0.022 & 0.500 \\
4 & 0.0000 & simple & mzi & mode\_expectations & bunched & 3/8 & MCC, F1, BalAcc & 0.483 & 0.022 & 0.500 \\
5 & 0.0000 & ring\_nn & mzi & mode\_correlations & pair\_bunched & 3/8 & MCC, F1, BalAcc & 0.479 & 0.022 & 0.501 \\
6 & 0.0100 & clements\_mesh & mzi & mode\_expectations & alternating & 2/8 & AUC, GPS & \textbf{0.576} & 0.000 & 0.500 \\
7 & 0.0100 & ring\_nn & mzi & probs\_fock & pair\_bunched & 2/8 & AUC, GPS & 0.573 & 0.000 & 0.500 \\
8 & 0.0100 & simple & bell & probs\_fock & pair\_bunched & 2/8 & AUC, GPS & 0.572 & 0.000 & 0.500 \\
9 & 0.0100 & clements\_mesh & mzi & mode\_expectations & bunched & 2/8 & AUC, GPS & 0.571 & 0.000 & 0.500 \\
10 & 0.0100 & clements\_mesh & mzi & mode\_correlations & bunched & 2/8 & AUC, GPS & 0.570 & 0.000 & 0.500 \\
\bottomrule
\end{tabular}%
}
\end{table*}

\begin{table*}[t]
\centering
\caption{Top-10 overlap table for NIFTY 50. Top10Count indicates in how many of 8 metrics a configuration appears in the top-10.}
\label{tab:top10_overlap_nifty50}
\setlength{\tabcolsep}{3pt}
\resizebox{.95\textwidth}{!}{%
\begin{tabular}{c l l l l l c l c c c}
\toprule
R & $\tau$ & Arch & Ent & Meas & State & Top10Count & MetricsHit & BestAUC & BestMCC & BestBalAcc \\
\midrule
1 & 0.0000 & ring\_nn & bell & mode\_expectations & pair\_bunched & 5/8 & MCC, BalAcc, AP, GPS, PSharpe & 0.511 & 0.021 & 0.507 \\
2 & 0.0000 & ring\_nn & mzi & mode\_expectations & pair\_bunched & 4/8 & MCC, BalAcc, GPS, PSharpe & 0.516 & 0.048 & 0.522 \\
3 & 0.0000 & ring\_nn & bell & mode\_expectations & bunched & 4/8 & MCC, BalAcc, GPS, PSharpe & 0.512 & 0.027 & 0.511 \\
4 & 0.0000 & clements\_mesh & bell & mode\_correlations & thermal\_like & 3/8 & AP, GPS, PSharpe & 0.520 & 0.003 & 0.500 \\
5 & 0.0000 & clements\_mesh & bell & mode\_correlations & pair\_bunched & 3/8 & AP, GPS, PSharpe & 0.516 & 0.002 & 0.500 \\
6 & 0.0000 & ring\_nn & mzi & probs\_fock & pair\_bunched & 3/8 & F1, AP, GPS & 0.514 & 0.000 & 0.500 \\
7 & 0.0000 & ring\_nn & mzi & mode\_correlations & bunched & 3/8 & MCC, BalAcc, AP & 0.512 & 0.051 & 0.511 \\
8 & 0.0000 & permuted\_ring & bell & mode\_correlations & bunched & 3/8 & MCC, F1, PSharpe & 0.500 & 0.024 & 0.501 \\
9 & 0.0000 & permuted\_ring & bell & mode\_expectations & alternating & 2/8 & F1, GPS & \textbf{0.521} & 0.000 & 0.500 \\
10 & 0.0000 & ring\_nn & bell & probs\_fock & pair\_bunched & 2/8 & AP, GPS & 0.514 & 0.000 & 0.500 \\
\bottomrule
\end{tabular}%
}
\end{table*}

\begin{table*}[t]
\centering
\caption{Top-10 overlap table for Crypto. Top10Count indicates in how many of 8 metrics a configuration appears in the top-10.}
\label{tab:top10_overlap_crypto}
\setlength{\tabcolsep}{3pt}
\resizebox{.95\textwidth}{!}{%
\begin{tabular}{c l l l l l c l c c c}
\toprule
R & $\tau$ & Arch & Ent & Meas & State & Top10Count & MetricsHit & BestAUC & BestMCC & BestBalAcc \\
\midrule
1 & 0.0000 & clements\_mesh & mzi & mode\_expectations & bunched & 3/8 & MCC, BalAcc, AP & 0.536 & 0.020 & 0.501 \\
2 & 0.0000 & clements\_mesh & mzi & mode\_expectations & alternating & 3/8 & MCC, BalAcc, AP & 0.536 & 0.031 & 0.503 \\
3 & 0.0000 & permuted\_ring & bell & mode\_expectations & thermal\_like & 3/8 & MCC, F1, BalAcc & 0.505 & 0.035 & 0.503 \\
4 & 0.0050 & clements\_mesh & bell & mode\_correlations & thermal\_like & 3/8 & MCC, BalAcc, PSharpe & 0.505 & 0.063 & 0.514 \\
5 & 0.0050 & clements\_mesh & mzi & mode\_correlations & thermal\_like & 3/8 & MCC, BalAcc, PSharpe & 0.503 & 0.042 & 0.509 \\
6 & 0.0000 & ring\_nn & bell & mode\_correlations & pair\_bunched & 3/8 & MCC, F1, BalAcc & 0.500 & 0.031 & 0.503 \\
7 & 0.0050 & clements\_mesh & bell & mode\_correlations & pair\_bunched & 3/8 & MCC, BalAcc, PSharpe & 0.493 & 0.063 & 0.505 \\
8 & 0.0100 & ring\_nn & mzi & mode\_correlations & thermal\_like & 2/8 & AUC, GPS & \textbf{0.552} & 0.000 & 0.500 \\
9 & 0.0100 & clements\_mesh & bell & probs\_fock & pair\_bunched & 2/8 & AUC, GPS & 0.550 & 0.000 & 0.500 \\
10 & 0.0100 & ring\_nn & mzi & probs\_fock & pair\_bunched & 2/8 & AUC, GPS & 0.546 & 0.000 & 0.500 \\
\bottomrule
\end{tabular}%
}
\end{table*}

\subsection{Comparison with Classical Baselines}

\begin{figure}[t]
  \centering
  \includegraphics[width=0.98\columnwidth]{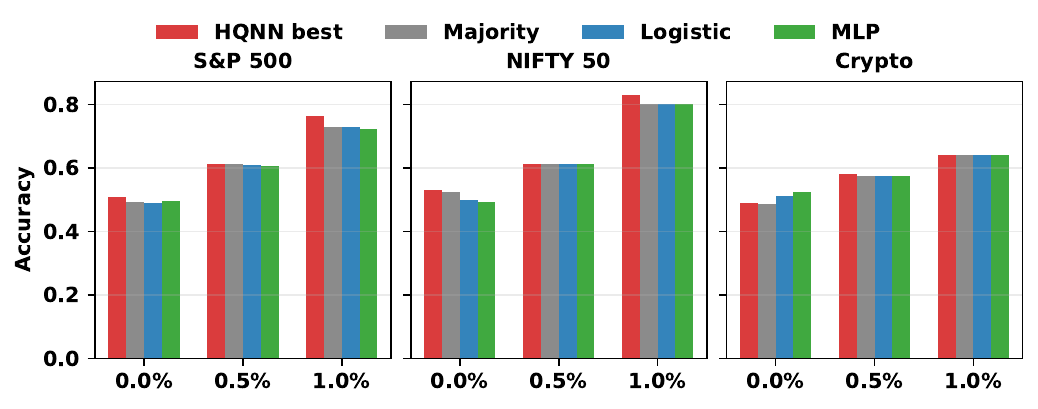}
  \caption{Accuracy comparison against majority, logistic, and MLP baselines.}
  \label{fig:baseline_comparison}
\end{figure}

Figure~\ref{fig:baseline_comparison} compares the best photonic HQNN identified by Q-PhotoMarket against three classical baselines: a majority-class classifier, logistic regression, and a shallow multilayer perceptron. Across the evaluated markets and return thresholds, the photonic HQNN consistently outperforms the majority baseline and achieves competitive performance relative to conventional machine learning models.

The performance gap becomes more apparent at higher return thresholds, where class imbalance increases and simple classifiers struggle to identify minority positive samples. Although the improvements over classical neural networks remain modest, these results demonstrate that systematic photonic architecture exploration enables HQNNs to remain competitive on realistic financial prediction tasks while providing valuable insights into photonic circuit design.




\subsection{Discussion}

The primary contribution of this work is not the identification of a single best-performing photonic HQNN, but the systematic characterization of the photonic design space. Across more than 5,000 evaluated configurations, the results consistently demonstrate the influence of architectural decisions on predictive performance. In particular, circuit topology and measurement strategy contribute substantially larger performance variations than input photon states or entangling primitives, suggesting that future photonic HQNN development should prioritize architectural exploration over isolated circuit optimization.

Among the investigated architectures, the Clements Mesh and, in several cases, the Simple and Ring NN topologies repeatedly appear among the highest-performing configurations. Their recurring presence across multiple evaluation metrics and financial markets indicates that expressive interferometer structures provide a favorable balance between representational capacity and trainability. Likewise, the influence of the measurement strategy highlights that the classical representation extracted from the photonic circuit is an integral component of the learning pipeline.

An interesting observation is that increasing the return threshold generally improves AUC despite introducing more severe class imbalance. By filtering out small price fluctuations, larger thresholds produce prediction tasks with clearer decision boundaries, enabling photonic HQNNs to better separate positive and negative samples. At the same time, this setting increases the risk of prediction collapse, emphasizing the importance of calibration-aware evaluation and imbalance-sensitive metrics.

This work has several limitations. All experiments are performed using differentiable photonic simulation, and the study focuses on daily market prediction using three representative financial datasets. In addition, the explored design space is limited to the investigated circuit topologies, measurement strategies, and encoding schemes. Future work will extend Q-PhotoMarket to larger photonic architectures, hardware-aware optimization, execution on real quantum devices, and additional real-world QML applications, while investigating automated architecture search techniques that further reduce the cost of exploring high-dimensional photonic design spaces.

\section{Conclusion}

This work presents Q-PhotoMarket, a systematic design space exploration framework for photonic hybrid quantum neural networks applied to financial market prediction. By jointly exploring multiple photonic design dimensions under a unified evaluation protocol, the proposed framework provides a methodology for understanding how architectural decisions influence learning performance. Q-PhotoMarket serves not only as a benchmark for photonic QML in finance but also as a foundation for future research on automated photonic architecture optimization and application-driven design space exploration across various QML domains.

\section*{Acknowledgment}
This work was supported in part by the NYUAD Center for Quantum and Topological Systems (CQTS), funded by Tamkeen under the NYUAD Research Institute grant CG008.

\bibliographystyle{IEEEtran}

\bibliography{refs}

\end{spacing}

\end{document}